\documentclass[onecolumn,floats,floatfix,nofootinbib,11pt]{revtex4-1}
\usepackage{graphicx}
\usepackage{dcolumn}
\usepackage{bm}
\usepackage{graphics}
\usepackage{slashed}
\usepackage{amssymb}
\usepackage{natbib}
\usepackage{amsmath}
\usepackage{amsthm}
\usepackage{url}
\usepackage[usenames,dvipsnames,svgnames,table]{xcolor}
\usepackage[T1]{fontenc}
\DeclareFontFamily{T1}{calligra}{}
\DeclareFontShape{T1}{calligra}{m}{n}{<->s*[1.44]callig15}{}
\DeclareMathAlphabet\mathcalligra   {T1}{calligra} {m} {n}
\DeclareMathAlphabet\mathzapf       {T1}{pzc} {mb} {it}
\DeclareMathAlphabet\mathchorus     {T1}{qzc} {m} {n}
\DeclareMathAlphabet\mathrsfso      {U}{rsfso}{m}{n}
\newcommand{\bea}{\begin{eqnarray}}
\newcommand{\ena}{\end{eqnarray}}
\newcommand{\bean}{\begin{eqnarray*}}
\newcommand{\enan}{\end{eqnarray*}}

\newtheorem{remark}{Remark}
\newtheorem{definition}{Definition}

\makeatletter
\newcommand{\tpitchfork}{%
  \vbox{
    \baselineskip\z@skip
    \lineskip-.52ex
    \lineskiplimit\maxdimen
    \m@th
    \ialign{##\crcr\hidewidth\smash{$-$}\hidewidth\crcr$\pitchfork$\crcr}
  }%
}
\makeatother

\newtheorem{prop}{Proposition}

\begin{document}
\title{Space and time ambiguities in vacuum electrodynamics}

\author{Érico Goulart} \email{egoulart@ufsj.edu.br}
\affiliation{Federal University of S\~ao Jo\~ao d'El Rei, C.A.P. Rod.: MG 443, KM 7, CEP-36420-000, Ouro Branco, MG, Brazil}
\author{Eduardo Bittencourt} \email{bittencourt@unifei.edu.br}
\affiliation{Federal University of Itajub\'a, Itajub\'a, Minas Gerais 37500-903, Brazil}

\date{\today}

\begin{abstract}
It is shown that every regular electromagnetic field in vacuum identically satisfy Maxwell equations in a new manifold where the roles of space and time have been exchanged. The new metric is Lorentzian, depends on the particular solution and forces the flow of time to tilt somehow in the direction of the field lines. We give a detailed description of the transformation and discuss several of its properties, both in the algebraic and differential settings. Examples are given where the new metrics are explicitly computed and carefully analyzed. We conclude with possible applications of the transformation as well as future perspectives.
\end{abstract}

\maketitle

\section{Introduction}

The conformal invariance of vacuum electrodynamics in four dimensions is known to exist since the early days of relativistic physics \cite{Cun,Bat}. Roughly speaking, the invariance means that all electromagnetic fields satisfying the source-free Maxwell equations in a space-time $(\mathcal{M}, \boldsymbol{g})$, also satisfy them in another space-time $(\mathcal{M}, \tilde{\boldsymbol{g}})$, with metric $\tilde{\boldsymbol{g}}=\Omega^{2}\boldsymbol{g}$, where $\Omega$ is a smooth and strictly positive function. As stressed in \cite{Harte}, this provides a sense in which electromagnetism alone cannot be used to measure certain aspects of geometry. In particular, since conformal transformations preserve the causal relationships and null geodesics, the conformal factor cannot be determined from pure electromagnetic experiments. Importantly, if the light cones of two Lorentz metrics $\boldsymbol{g}$ and $\tilde{\boldsymbol{g}}$ coincide at a point $p\in\mathcal{M}$, then at $p$, $\tilde{\boldsymbol{g}}$ must be a multiple of $\boldsymbol{g}$ (see, for instance, \cite{Wald,H-E}).

It is well-known that if two non-degenerate metrics of arbitrary signature yield the same Hodge dual for all 2-forms in four dimensions, then they must be conformally related \cite{Dray}. Little-known, however, is that many more metrics are allowed if we restrict ourselves to the invariance of the Hodge dual for a particular 2-form. This was first established in \cite{GT}, where they showed that any \textit{particular} solution of the source-free Maxwell's equations compatible with a metric $\boldsymbol{g}$ is also compatible with the \textit{disformal metric}
\begin{equation}\label{firstdef}
\boldsymbol{g}\mapsto\tilde{\boldsymbol{g}}=A\boldsymbol{g}+B\boldsymbol{\Phi},
\end{equation}
where $A,B\in\mathbb{R}$ and $\boldsymbol{\Phi}\in\mbox{Sym}^{0}_{2}$ depend on the electromagnetic 2-form in a specific way. In contrast with conformal transformations, the disformal transformations do not preserve angles and, hence, the causal structure of the underlying space-time. Recently, it has been found \cite{Harte} a much larger class of metric transformations—involving five free functions—which preserve the Maxwell solutions both in vacuum, without local currents, and also for the force-free electrodynamics associated with a tenuous plasma.

More generally, disformally-related metrics appear in a diversity of situations of physical relevance. In general relativity, for instance, Kerr-Schild metrics provide the simpler examples of solutions of Einstein's equations for which $g_{ab}=\eta_{ab}$ and the disformal field has the form $\Phi_{ab}=l_{a}l_{b}$, with $l^{a}l_{a}=0$. Examples of the latter include the Schwarzschild, Kerr and plane wave metrics \cite{Kerr}. Similar deformations appear also in the research programme of analogue gravity \cite{AG, ABH}, which investigates analogues of general relativistic gravitational fields within other physical systems. In the last few years, disformal transformations have been used as a simple modification of the space-time metric that can help in different approaches: quantum gravity phenomenology \cite{magueijo04,clb16,amelino01}, MOND \cite{beken_mond}, scalar-tensor theories of gravitation \cite{scalartheory,mota,ip,sak1,sak2}, Mimetic gravity \cite{rua,matarrese,sunny1,sunny2} and Horndeski theory \cite{miguel1,vernizzi,dario}, with applications in field theory \cite{uzan,yuan,brax1,brax2,nov_bit_gordon,nov_bit_drag}, particle physics \cite{bitt_nov_faci,nov_bit,erico12,bitt15} and the singularity theorems \cite{bitt20}.

In this paper we revisit the disformal invariance of Maxwell's equations in vacuum, as presented in \cite{GT}, and discuss several of its geometrical/topological consequences not described before. Our main result is that the disformal transformation somehow implies in an ambiguity between space and time. In other words: in the \textit{disformal manifold} $(\mathcal{M},\tilde{\boldsymbol{g}})$, time does not flow inside the physical light cone, but rather in a direction specified by the field lines. More specifically, restricting our attention to regular 2-forms, we show that: i) the disformal metric is proportional to the physical energy-momentum tensor; ii) conversely, the disformal energy-momentum tensor is proportional to the physical metric; iii) conformal and disformal transformations commute; iv) electric and magnetic 3-vectors are time-like in $(\mathcal{M},\tilde{\boldsymbol{g}})$; v) only the anisotropic pressure transforms for a specific observer; vi) Newman-Penrose scalars transform in an unexpected way. We give some examples for which the disformal metric is explicitly calculated and the disformal causal structure scrutinized. We hope that our results may shed light in obtaining new solutions in flat or curved space-times endowed with exotic causal structures. Throughout, we use units such that $c=1$ and $8\pi G=1$.

\section{Mathematical machinery}

 \subsection{Electromagnetic 2-forms at a space-time point}

In this section we briefly recall some algebraic properties the of electromagnetic fields. To begin with, we let the \textit{physical manifold} $(\mathcal{M},\boldsymbol{g})$ denote a four-dimensional, oriented, space-time with metric signature $(1,3)$. At a space-time point, the electromagnetic 2-form and its Hodge dual induced by $\boldsymbol{g}$ are expanded as\footnote{We refer the reader to the Appendix (\ref{notation}) for further details on our conventions.}
\begin{equation}
\boldsymbol{F}=\frac{1}{2}F_{ab}dx^{a}\wedge dx^{b},\quad\quad\quad \star_{g}\boldsymbol{F}=\frac{1}{2}\star_{g}F_{ab}dx^{a}\wedge dx^{b}.
\end{equation}
Given $\boldsymbol{F}\in\Lambda^{2}(T_{p}^{*}\mathcal{M})$ and an arbitrary vector $\boldsymbol{X}\in T_{p}\mathcal{M}$, a direct calculation gives the identity
\begin{equation}\label{Gen}
\star_{g}[(i_{\boldsymbol{X}}\star_{g}\boldsymbol{F})\wedge \boldsymbol{X}^{\flat_{g}}]=||\boldsymbol{X}||_{g}^{2}\boldsymbol{F}+(i_{\boldsymbol{X}}\boldsymbol{F})\wedge \boldsymbol{X}^{\flat_{g}}.
\end{equation}
In the particular case of a future-directed, normalized, time-like vector i.e., $||\boldsymbol{X}||^{2}_{g}=1$, there follows the (3+1) decomposition
\begin{equation}\nonumber
\boldsymbol{F}=\boldsymbol{E}^{\flat_{g}}\wedge \boldsymbol{X}^{\flat_{g}}+\star_{g}(\boldsymbol{B}^{\flat_{g}}\wedge \boldsymbol{X}^{\flat_{g}}),
\end{equation}
with the \textit{projections} defined by
\begin{equation}
\boldsymbol{E}^{\flat_{g}}\equiv -i_{\boldsymbol{X}}\boldsymbol{F},\quad\quad\quad \boldsymbol{B}^{\flat_{g}}\equiv i_{\boldsymbol{X}}\star_{g}\boldsymbol{F}.
\end{equation}
The space-like quantities $\boldsymbol{E},\boldsymbol{B}\in \boldsymbol{X}^{\perp_{g}}$ are interpreted as the electric and magnetic $3$-vectors, respectively, as measured by the \textit{observer} $\boldsymbol{X}$ at the point $p$.  Henceforth, for the sake of simplicity, their squared norms are written as
 \begin{equation}
 E^{2}\equiv-||\boldsymbol{E}||_{g}^{2},\quad\quad\quad B^{2}\equiv-||\boldsymbol{B}||_{g}^{2}.
 \end{equation}

With these conventions, the invariant scalars of the electromagnetic 2-form are given by
\begin{eqnarray}\label{invariants}
&&\phi_{g}\equiv \langle \boldsymbol{F},\star_{g}\boldsymbol{F}\rangle_{g}=2EB\mbox{cos}\ \theta,\quad\quad\quad \psi_{g}\equiv \langle \boldsymbol{F},\boldsymbol{F}\rangle_{g}=B^{2}-E^{2},
\end{eqnarray}
where $\theta$ is the angle between the electric and the magnetic 3-vectors. From the latter we construct a third invariant
\begin{equation}
\kappa_{g}\equiv\frac{1}{2}\sqrt{\psi_{g}^{2}+\phi_{g}^{2}}=\frac{1}{2}\sqrt{E^{4}+B^{4}+2E^{2}B^{2}\mbox{cos}\ 2\theta},
\end{equation}
and then, the electromagnetic 2-form is called \textit{null} if $\kappa_{g}=0$ or \textit{regular} if $\kappa_{g}\neq 0$. Notice that $\kappa_{g}$ vanishes iff the electric and magnetic magnitudes are equal and the corresponding 3-vectors are perpendicular to one another. Conversely, if $\kappa_{g}\neq 0$ there exists an observer $\boldsymbol{X}$ such that the electric and magnetic 3-vectors lie on the same line in $\boldsymbol{X}^{\perp_{g}}$. Their directions may be the same, opposite or one of the 3-vectors vanishes. This configuration is called an \textit{electromagnetic wrench} and, in this case, one has the relations
\begin{equation}
\boldsymbol{E}=\alpha\boldsymbol{W},\quad\quad\quad  \boldsymbol{B}=\beta\boldsymbol{W},
\end{equation}
for scalars $\alpha,\beta\in\mathbb{R}$ and $\boldsymbol{W}\in\boldsymbol{X}^{\perp_{g}}$ with $||\boldsymbol{W}||_{g}^{2}=-1$. Since $\boldsymbol{W}$ is defined up to a sign, we shall assume
\begin{equation}\label{assumptions}
\alpha> 0\quad\mbox{if}\quad\psi_{g}\leq 0,\quad\quad\quad \beta> 0\quad\mbox{if}\quad\psi_{g}> 0.
\end{equation}
We call such an $\boldsymbol{X}$ by a \textit{wrench observer} and $\boldsymbol{W}$ by a \textit{wrench 3-vector}. If there are light rays one directed along $\boldsymbol{W}$ and another in the opposite direction, the corresponding rays determine two null vectors $\boldsymbol{l}$ and $\boldsymbol{n}$, respectively. They constitute the real principal null directions of $\boldsymbol{F}$ and there follow that every possible pair $\{\boldsymbol{X},\boldsymbol{W}\}$ lies on the plane spanned by them \cite{synge, wheeler}.

It is well known that every electromagnetic 2-form satisfies the quadratic algebraic identities
\begin{equation}\label{Ruse3}
\star_{g}\boldsymbol{F}^{2_{g}}-\boldsymbol{F}^{2_{g}}=\psi_{g}\boldsymbol{g},\quad\quad\quad (\star_{g}\boldsymbol{F}\boldsymbol{F})_{g}=-\frac{1}{2}\phi_{g}\boldsymbol{g},
\end{equation}
and from the latter one obtains the higher order relations
\begin{eqnarray}\label{cubic}
&&\boldsymbol{F}^{3_{g}}+\psi_{g}\boldsymbol{F}+\frac{1}{2}\phi_{g}\star_{g}\boldsymbol{F}=0,\\\label{quartic}
&&\boldsymbol{F}^{4_{g}}+\psi_{g}\boldsymbol{F}^{2_{g}}-\frac{1}{4}\phi_{g}^{2}\boldsymbol{g}=0.
\end{eqnarray}
Notice that the last equation is also a consequence of the Cayley-Hamilton theorem and, since null 2-forms are highly degenerated from the algebraic point of view, we shall deal only with regular 2-forms in what follows (see \cite{GT} for some results concerning the null case).

\subsection{Energy-momentum tensor}

\quad\ Given the electromagnetic 2-form and the metric at a space-time point, one constructs the \textit{energy-momentum tensor} as follows
\begin{equation}\label{defT}
\boldsymbol{T}\equiv\boldsymbol{F}^{2_{g}}+\frac{\psi_{g}}{2}\boldsymbol{g},\qquad{\rm with}\qquad\mbox{tr}_{g}(\boldsymbol{T})=0.
\end{equation}
Several algebraic relations may then be obtained from Eqs.\ (\ref{Ruse3}), (\ref{cubic}) and (\ref{quartic}). In particular, one shows that
\begin{equation}\label{Sq}
\boldsymbol{T}^{2_{g}}=\kappa_{g}^{2}\boldsymbol{g},\quad\quad\quad \mbox{tr}_{g}(\boldsymbol{T}^{2_{g}})=4\kappa_{g}^{2}.
\end{equation}
This equation was first established by Ruse in \cite{Ruse} and simple manipulations gives the determinantal relation
\begin{equation}\label{detT}
\sqrt{-T}=\kappa_{g}^{2}\sqrt{-g}\neq 0,
\end{equation}
with $T\equiv\mbox{det}(T_{ab})$. This means that the energy-momentum tensor of a regular 2-form is invertible, in contrast with the case of a null 2-form where the above determinant identically vanishes. Less known relations involving $\boldsymbol{F}$ and $\boldsymbol{T}$ are the following
\begin{eqnarray}\label{TF}
&&(\boldsymbol{T}\boldsymbol{F})_{g}=-\frac{1}{2}(\psi_{g}\boldsymbol{F}+\phi_{g}\star_{g}\boldsymbol{F}),\\\label{TFT}
&&(\boldsymbol{T}\boldsymbol{F}\boldsymbol{T})_{g}=\kappa_{g}^{2}\boldsymbol{F},\\\label{FTF}
&&(\boldsymbol{F}\boldsymbol{T}\boldsymbol{F})_{g}=-\frac{\psi_{g}}{2}\boldsymbol{T}+\kappa_{g}^{2}\boldsymbol{g}.
\end{eqnarray}
Notice that Eqs.\ (\ref{TF}) and (\ref{TFT}) define new 2-forms whereas Eq.\ (\ref{FTF}) yields a symmetric tensor. Again, the right hand sides of these equation would be trivial in the case of null 2-forms.

Now, a generic observer $\boldsymbol{X}\in T_{p}\mathcal{M}$ decomposes the energy-momentum tensor as follows
\begin{equation}\label{T}
\boldsymbol{T}=\rho\,\boldsymbol{X}^{\flat_{g}}\otimes\boldsymbol{X}^{\flat_{g}}-p\,\boldsymbol{h}+\boldsymbol{q}^{\flat_{g}}\otimes\boldsymbol{X}^{\flat_{g}}+\boldsymbol{X}^{\flat_{g}}\otimes\boldsymbol{q}^{\flat_{g}}+\boldsymbol{\Pi},
\end{equation}
where the energy-density, pressure, heat-flux and anisotropic pressure are, respectively, given by
\begin{eqnarray}\label{rho}
&&\rho=\frac{1}{2}(E^{2}+B^{2}),\quad\quad p=\frac{1}{6}(E^{2}+B^{2}),\quad\quad\boldsymbol{q}^{\flat_{g}}=-i_{\boldsymbol{X}}*_{g}(\boldsymbol{E}^{\flat_{g}}\wedge\boldsymbol{B}^{\flat_{g}}),\\\label{pi}
&&\quad\quad\quad\quad\boldsymbol{\Pi}=-\boldsymbol{E}^{\flat_{g}}\otimes\boldsymbol{E}^{\flat_{g}}-\boldsymbol{B}^{\flat_{g}}\otimes\boldsymbol{B}^{\flat_{g}}-\frac{1}{3}(E^{2}+B^{2})\boldsymbol{h},
\end{eqnarray}
with $\boldsymbol{h}\equiv\boldsymbol{g}-\boldsymbol{X}^{\flat_{g}}\otimes\boldsymbol{X}^{\flat_{g}}$ the projector induced by $\boldsymbol{X}$. For a wrench observer, however, the above decomposition drastically simplifies. Since $\boldsymbol{E}$ and $\boldsymbol{B}$ are parallel, the heat flux identically vanishes and one has the equality $\rho=\kappa_{g}$. In this case, Eq.\ (\ref{T}) reduces to
\begin{equation}\label{TXW}
\boldsymbol{T}=\kappa_{g}\left[2(\boldsymbol{X}^{\flat_{g}}\otimes\boldsymbol{X}^{\flat_{g}}-\boldsymbol{W}^{\flat_{g}}\otimes\boldsymbol{W}^{\flat_{g}})-\boldsymbol{g}\right].
\end{equation}

Here, it is important to notice that the signature of $\boldsymbol{T}$ is $(3,1)$. In other words, the energy-momentum tensor associated to a regular electromagnetic 2-form is a tensor of Lorentzian type, though its signature is the opposite of that of $\boldsymbol{g}$. Furthermore, it turns out that $\kappa_{g}$ may be interpreted as the minimum energy-density among all possible energy-densities observed by different observers.

\section{Disformal transformations}

\subsection{Algebraic setting}

The fact that $\boldsymbol{T}$ is an invertible tensor with Lorentzian signature motivates the following
\begin{definition}
A disformal transformation of the metric $\boldsymbol{g}$ induced by a regular 2-form $\boldsymbol{F}$ is the mapping
\begin{equation}\label{disf}
(\boldsymbol{F},\boldsymbol{g})\mapsto\tilde{\boldsymbol{g}}\equiv -\kappa^{-1}_{g}\left(\boldsymbol{F}^{2_{g}}+\frac{\psi_{g}}{2}\boldsymbol{g}\right)=-\kappa^{-1}_{g}\boldsymbol{T}.
\end{equation}
\end{definition}
In a coordinate basis at $p$, we have
\begin{equation}
\tilde{g}_{ab}= -\kappa_{g}^{-1}T_{ab},\quad\quad\quad (\tilde{g}^{-1})^{ab}= -\kappa_{g}^{-1}(g^{-1})^{ac}(g^{-1})^{bd}T_{cd},
\end{equation}
where the second relation stems from Eq.\ (\ref{Ruse3}). Clearly, this is a volume-preserving map since a direct calculation gives $|\tilde{g}|=|g|$. This is in contrast with conformal transformations where angles are preserved but scales are modified.
\begin{remark}
Notice that $\tilde{\boldsymbol{g}}$ has signature $(1,3)$ and carries the same physical dimensions of $\boldsymbol{g}$. It turns out that $\tilde{\boldsymbol{g}}$ is the unique symmetric rank-2 tensor constructed solely with the metric and the electromagnetic 2-form satisfying the above statements.
\end{remark}

Let us discuss some geometrical properties induced by the disformal transformation. First, consider the inner product induced by $\tilde{\boldsymbol{g}}$ on $\Lambda^{2}(T_{p}^{*}\mathcal{M})$. For arbitrary 2-forms $\boldsymbol{G},\boldsymbol{H}$, there follows
\begin{equation}\label{Innertilde}
\langle \boldsymbol{G},\boldsymbol{H}\rangle_{\tilde{g}}= -\frac{1}{2}\kappa_{g}^{-2}\mbox{tr}_{g}(\boldsymbol{G}\boldsymbol{T}\boldsymbol{H}\boldsymbol{T})\neq\langle \boldsymbol{G},\boldsymbol{H}\rangle_{g}.
\end{equation}
However, letting $\boldsymbol{H}=\boldsymbol{F}$ and using Eq.\ (\ref{TFT}), we obtain
\begin{equation}\label{1}
\langle \boldsymbol{F},\boldsymbol{G}\rangle_{\tilde{g}}=\langle \boldsymbol{F},\boldsymbol{G}\rangle_{g},
\end{equation}
meaning that both metrics define the same inner products as far as one of the factors is precisely the electromagnetic 2-form. What about the Hodge endomorphism associated to $\tilde{\boldsymbol{g}}$? We have
\begin{equation}
\star_{\tilde{g}}:\Lambda^{2}(T_{p}^{*}\mathcal{M})\rightarrow\Lambda^{2}(T_{p}^{*}\mathcal{M}),\quad\quad\quad\boldsymbol{G}\mapsto\star_{\tilde{g}}\boldsymbol{G},
\end{equation}
with
\begin{eqnarray*}\label{tilde*}
\star_{\tilde{g}}G_{ab}&=&\frac{1}{2}\sqrt{-\tilde{g}}[abpq](\tilde{g}^{-1})^{pc}(\tilde{g}^{-1})^{qd}G_{cd}\\
&=&\frac{1}{2}\sqrt{-g}[abpq](\tilde{g}^{-1})^{pc}(\tilde{g}^{-1})^{qd}G_{cd}\neq\star_{g}G_{ab}.
\end{eqnarray*}
Again, putting $G_{ab}=F_{ab}$ and using Eq.\ (\ref{TFT}) gives the important identity
\begin{equation}\label{2}
\star_{\tilde{g}}\boldsymbol{F}= \star_{g}\boldsymbol{F}.
\end{equation}
In other words, the disformal transformation preserve the Hodge duality of the electromagnetic 2-form. Finally, combining Eqs.\ (\ref{1}) and (\ref{2}), one easily shows that
\begin{equation}\label{invdisf}
\psi_{\tilde{g}}\equiv\langle \boldsymbol{F},\boldsymbol{F}\rangle_{\tilde{g}}=\psi_{g},\quad\quad\quad \phi_{\tilde{g}}=\langle \boldsymbol{F},*_{\tilde{g}}\boldsymbol{F}\rangle_{\tilde{g}}=\phi_{g},
\end{equation}
i.e., the invariants of the electromagnetic 2-form are kept intact by the disformal mapping. In particular, one obtains $\kappa_{\tilde{g}}=\kappa_{g}$.

\begin{remark}
Due to the invariance of the scalars, the disformal transformation maps regular electromagnetic 2-forms into regular electromagnetic 2-forms. Particularly, an electrically/magnetically dominated configuration is transformed into a new electrically/magnetically configuration.
\end{remark}

We now consider the energy-momentum tensor constructed with $\boldsymbol{F}$ and $\tilde{\boldsymbol{g}}$. Starting from Eq.\ (\ref{defT}), substituting $\tilde{\boldsymbol{g}}$ by $\boldsymbol{g}$, and using Eq. (\ref{disf}), we have:
\begin{equation}
\tilde{\boldsymbol{T}}=\boldsymbol{F}^{2_{\tilde{g}}}+\frac{\psi_{\tilde{g}}}{2}\tilde{\boldsymbol{g}}=-\kappa^{-1}_{g}(\boldsymbol{F}\boldsymbol{T}\boldsymbol{F})_{g}+\frac{\psi_{\tilde{g}}}{2}\tilde{\boldsymbol{g}}.
\end{equation}
Using Eqs.\ (\ref{FTF}) and (\ref{invdisf}), we obtain, after simple algebra
\begin{equation}\label{Ttilde}
\tilde{\boldsymbol{T}}=-\kappa_{g}\boldsymbol{g},
\end{equation}
showing that $\tilde{\boldsymbol{T}}$ is nothing but a conformal transformation of the background metric. Therefore, one concludes that the disformal mapping is an unusual type of transformation that somehow interchanges geometry with energy-momentum and vice-versa.

In order to further clarify the transformation described above, consider a set of complex null tetrad $\{\boldsymbol{l}, \boldsymbol{n}, \boldsymbol{m}, \bar{\boldsymbol{m}}\}$ satisfying the normalization/orientation conditions described in Appendix (\ref{np-appendix}). In general, one has the following complex nonvanishing scalars associated to the electromagnetic 2-form
\begin{equation}
\Phi_{0}=l^{a}m^{b}F_{ab},\quad\quad\quad\Phi_{1}=\frac{1}{2}(l^{a}n^{b}+\bar{m}^{a}m^{b})F_{ab},\quad\quad\quad\Phi_{2}=\bar{m}^{a}n^{b}F_{ab}.
\end{equation}
However, taking the basis aligned with the principal null directions of $\boldsymbol{F}$ as defined before, only $\Phi_{1}$ survives. This is immediate, since when $F^{a}_{\phantom a b}$ is applied to a vector in the basis it returns another vector proportional to the former. In this case we may verify that (see, for instance, \cite{Frolov})
\begin{equation}
\frac{1}{2}\boldsymbol{F}=-\mbox{Re} (\Phi_{1})\ \boldsymbol{l}^{\flat_{g}}\wedge\boldsymbol{n}^{\flat_{g}}+i\,\mbox{Im} (\Phi_{1})\ \boldsymbol{m}^{\flat_{g}}\wedge\bar{\boldsymbol{m}}^{\flat_{g}}.
\end{equation}
Within this adapted frame, the background metric and energy-momentum tensor are expanded as
\begin{eqnarray}
\boldsymbol{g}&=&\boldsymbol{l}^{\flat_{g}}\otimes\boldsymbol{n}^{\flat_{g}}+\boldsymbol{n}^{\flat_{g}}\otimes\boldsymbol{l}^{\flat_{g}}-\boldsymbol{m}^{\flat_{g}}\otimes\bar{\boldsymbol{m}}^{\flat_{g}}-\bar{\boldsymbol{m}}^{\flat_{g}}\otimes\boldsymbol{m}^{\flat_{g}},\label{g_null_basis}\\
\boldsymbol{T}&=&\kappa_{g}\left(\boldsymbol{l}^{\flat_{g}}\otimes\boldsymbol{n}^{\flat_{g}}+\boldsymbol{n}^{\flat_{g}}\otimes\boldsymbol{l}^{\flat_{g}}+\boldsymbol{m}^{\flat_{g}}\otimes\bar{\boldsymbol{m}}^{\flat_{g}}+\bar{\boldsymbol{m}}^{\flat_{g}}\otimes\boldsymbol{m}^{\flat_{g}}\right).
\end{eqnarray}
Similarly, the corresponding disformal quantities are given by
\begin{eqnarray}
\tilde{\boldsymbol{g}}&=&-\boldsymbol{l}^{\flat_{g}}\otimes\boldsymbol{n}^{\flat_{g}}-\boldsymbol{n}^{\flat_{g}}\otimes\boldsymbol{l}^{\flat_{g}}-\boldsymbol{m}^{\flat_{g}}\otimes\bar{\boldsymbol{m}}^{\flat_{g}}-\bar{\boldsymbol{m}}^{\flat_{g}}\otimes\boldsymbol{m}^{\flat_{g}},\label{tilde_g_null_basis}\\
\tilde{\boldsymbol{T}}&=&\kappa_{g}\left(-\boldsymbol{l}^{\flat_{g}}\otimes\boldsymbol{n}^{\flat_{g}}-\boldsymbol{n}^{\flat_{g}}\otimes\boldsymbol{l}^{\flat_{g}}+\boldsymbol{m}^{\flat_{g}}\otimes\bar{\boldsymbol{m}}^{\flat_{g}}+\bar{\boldsymbol{m}}^{\flat_{g}}\otimes\boldsymbol{m}^{\flat_{g}}\right).
\end{eqnarray}
Notice that the following relations must hold
\begin{equation}
\boldsymbol{l}=\frac{1}{\sqrt{2}}(\boldsymbol{X}+\boldsymbol{W}),\quad\quad\quad \boldsymbol{n}=\frac{1}{\sqrt{2}}(\boldsymbol{X}-\boldsymbol{W}),\quad\quad\quad\kappa_{g}=2\Phi_{1}\bar{\Phi}_{1}.
\end{equation}
with $\boldsymbol{X}$ being a wrench observer and $\boldsymbol{W}$ the corresponding wrench 3-vector. From now on, all complex null tetrads are assumed to be adapted in this way. We shall see that the above decomposition, when suitably modified to the differential setting, facilitates the geometrical interpretation of the disformal metric.

\subsection{Differential setting}

The discussion presented so far may be easily generalized to the differential setting. To do so, we define the bundle of alternating $2$-forms by
\begin{equation}
\Lambda^{2}T^{*}\mathcal{M}=\bigcup\limits_{p\in \mathcal{M}}\Lambda^{2}(T_{p}^{*}\mathcal{M})
\end{equation}
and denote the space of smooth sections by
\begin{equation}
\Omega^{2}(\mathcal{M})=\Gamma(\Lambda^{2}T^{*}\mathcal{M}).
\end{equation}
Algebraic operations such as wedge products, interior products, Hodge duals and disformal transformations are then defined pointwise. We have the following immediate results:
\begin{prop}
If a pair of smooth 2-forms $\boldsymbol{F},\star_{g}\boldsymbol{F}\in\Omega^{2}(\mathcal{M})$ is closed under exterior derivation, then the pair $\boldsymbol{F},\star_{\tilde{g}}\boldsymbol{F}\in\Omega^{2}(\mathcal{M})$ is also closed.
\end{prop}
\begin{proof}
Since $\star_{\tilde{g}}\boldsymbol{F}$ and $\star_{g}\boldsymbol{F}$ coincide and exterior derivatives do not involve metric considerations, closedness follows trivially. In other words, if $\boldsymbol{F}$ is a solution of the source-free Maxwell equations in $(\mathcal{M},\boldsymbol{g})$ i.e.
\begin{equation}
\boldsymbol{d}\boldsymbol{F}=0,\quad\quad\quad \boldsymbol{d}\star_{g}\boldsymbol{F}=0,
\end{equation}
it identically satisfies them in $(\mathcal{M},\tilde{\boldsymbol{g}})$. In particular, letting $\tilde{\nabla}$ denote the covariant derivative operator compatible with the disformal metric $\tilde{\textbf{g}}$, there follow:
\begin{equation}\label{disfMax}
(\tilde{g}^{-1})^{ac}\tilde{\nabla}_{a}F_{bc}=0,\quad\quad\quad (\tilde{g}^{-1})^{ac}\tilde{\nabla}_{a}(\star_{\tilde{g}}F_{bc})=0.
\end{equation}
Notice, however, that $\tilde{\nabla}$ depends on $\boldsymbol{g}$ and on the particular solution $\boldsymbol{F}$ in a nontrivial, nonlinear, fashion. In contrast with the conformal transformations, where all solutions are mapped into new solutions, the disformal transformations somehow entangles the particular solution with a particular disformal metric.
\end{proof}

\begin{prop}
The quantity $\tilde{\boldsymbol{T}}$, as given by Eq.\ (\ref{Ttilde}), is identically conserved in $(\mathcal{M},\tilde{\boldsymbol{g}})$.
\end{prop}
\begin{proof}
The proof is entirely analogous to showing the conservation of $\boldsymbol{T}$ in $(\mathcal{M},\boldsymbol{g})$ i.e., starting from the definition
\begin{equation}
(\tilde{g}^{-1})^{ac}\tilde{\nabla}_{a}\tilde{T}_{bc}=(\tilde{g}^{-1})^{ac}\tilde{\nabla}_{a}\left(F_{bd}(\tilde{g}^{-1})^{de}F_{ec}+\frac{\tilde{\psi}}{2}\tilde{g}_{bc}\right),
\end{equation}
and using $\tilde{\nabla}_{a}\tilde{g}_{bc}=0$ one obtains, after applying Maxwell's equations in the form of Eq.\ (\ref{disfMax}), the vanishing of the covariant divergence
\begin{equation}
(\tilde{g}^{-1})^{ac}\tilde{\nabla}_{a}\tilde{T}_{bc}=0\quad\rightarrow\quad\tilde{\nabla}\cdot\tilde{\boldsymbol{T}}=0.
\end{equation}
\end{proof}
\begin{prop}
The composition of conformal and disformal transformations commute.
\end{prop}
\begin{proof}
The proof is straightforward.
\end{proof}

\section{The geometry $(\mathcal{M},\tilde{\boldsymbol{g}})$}

In this section we briefly discuss some geometrical aspects induced by the disformal metric on $\mathcal{M}$. In particular, we show how the causal structure is modified, the 3-vectors and hydrodynamic quantities transform and what are the interconnections between the Newman-Penrose (NP) scalars.

\subsection{Causality}

The fact that $\tilde{\boldsymbol{g}}$ carries a signature $(1,3)$ by no means implies that its causal structure coincide with the causal structure of the background metric. Indeed, start with an arbitrary solution of Maxwell's equations in vacuum and consider the disformal metric given by Eq.\ (\ref{disf}). Using the decomposition given by Eq.\ (\ref{TXW}), one shows that there exists a pseudo-orthonormal tetrad field $\{{\boldsymbol{e}}_{\mu}\}$, such that
\begin{equation}
\boldsymbol{g}(\boldsymbol{e}_{\mu},\boldsymbol{e}_{\nu})=\mbox{diag}(+1,-1,-1,-1),\quad\quad\quad\boldsymbol{\tilde{\boldsymbol{g}}}(\boldsymbol{e}_{\mu},\boldsymbol{e}_{\nu})=\mbox{diag}(-1,+1,-1,-1),
\end{equation}
Clearly, this basis is such that $\{\boldsymbol{e}_{0}, \boldsymbol{e}_{1}\}$ coincides with an appropriate wrench pair $\{\boldsymbol{X},\boldsymbol{W}\}$ for the field, with $\{\boldsymbol{e}_{1},\boldsymbol{e}_{2},\boldsymbol{e}_{3}\}$ forming a right-handed triad in $\boldsymbol{X}^{\perp_{g}}$.

Now, letting $\mathcal{C}$ and $\tilde{\mathcal{C}}$ denote the null cones with respect to $\boldsymbol{g}$ and $\tilde{\boldsymbol{g}}$, one sees that they drastically differ at all points: the \textit{disformal cones} twist in the direction of the line defined by $\boldsymbol{W}$ in an unexpected way. Indeed, at every space-time point, one may obtain the disformal cone by simply rotating (without deforming) the background cone through an angle of $\pi/2$ in the 2-plane spanned by the real null principal directions. Obviously, both cones share the directions of $\boldsymbol{l}$ and $\boldsymbol{n}$. Consequently, in spite of the fact that the disformal metric is Lorentzian, the roles of $\boldsymbol{e}_{0}$ and $\boldsymbol{e}_{1}$ are somehow flipped: $\boldsymbol{e}_{1}$ becomes time-like whereas $\boldsymbol{e}_{0}$ becomes space-like.

We then have, at a space-time point, two equivalence classes containing all time-like tangent vectors with respect to $\tilde{\boldsymbol{g}}$. We can arbitrarily call one of these equivalence classes ``future-directed'' and call the other ``past-directed''. Physically, this designation corresponds to a choice of an arrow of time at the point and we choose the future in $(\mathcal{M},\tilde{\boldsymbol{g}})$ as given by the direction of a wrench 3-vector $\boldsymbol{W}$ at $p$. This means that we can read off the \textit{disformal future} by looking at the orientation of field lines as defined by Eqs.\ (\ref{assumptions}). Interestingly, all static solutions in $(\mathcal{M},\boldsymbol{g})$ become time-dependent in $(\mathcal{M},\tilde{\boldsymbol{g}})$. Furthermore, we shall see in the next section that this natural definition may break the global hyperbolicity and may lead to closed time-like curves in the disformal manifold.

\subsection{3-vectors and hydrodynamic quantities}

How do the electric and magnetic 3-vectors appear in $(\mathcal{M},\tilde{\boldsymbol{g}})$? We proceed as follows: for any vector field $\tilde{\boldsymbol{X}}\in\mathfrak{X}(M)$, satisfying the normalization condition $||\tilde{\boldsymbol{X}}||_{\tilde{g}}=1$, we have the natural definitions
\begin{equation}\label{EBtilde}
\tilde{\boldsymbol{E}}^{\flat_{\tilde{g}}}\equiv -i_{\tilde{\boldsymbol{X}}}\boldsymbol{F},\quad\quad\quad \tilde{\boldsymbol{B}}^{\flat_{\tilde{g}}}\equiv i_{\tilde{\boldsymbol{X}}}\star_{\tilde{g}}\boldsymbol{F}.
\end{equation}
Now, if $\boldsymbol{X}\in\mathfrak{X}(M)$ is a congruence of wrench observers in $(\mathcal{M},\boldsymbol{g})$ and $\boldsymbol{W}\in\mathfrak{X}(M)$ is the corresponding wrench 3-vector field, we have the decomposition
\begin{equation}\label{rstilde}
\boldsymbol{F}=\alpha\boldsymbol{W}^{\flat_{g}}\wedge\boldsymbol{X}^{\flat_{g}}+\beta\star_{g}(\boldsymbol{W}^{\flat_{g}}\wedge\boldsymbol{X}^{\flat_{g}}),
\end{equation}
for real scalars $\alpha,\beta$ as before. Substituting Eq.\ (\ref{rstilde}) into Eq.\ (\ref{EBtilde}) and taking $\tilde{\boldsymbol{X}}=\boldsymbol{W}$, one obtains, after simple manipulations
\begin{equation}
\tilde{\boldsymbol{E}}^{\flat_{\tilde{g}}}=\alpha\boldsymbol{X}^{\flat_{g}},\quad\quad\quad \tilde{\boldsymbol{B}}^{\flat_{\tilde{g}}}=\beta\boldsymbol{X}^{\flat_{g}},
\end{equation}
since the Hodge duals coincide. As the 1-forms are proportional, one concludes that the wrench 3-vector in $(\mathcal{M},\boldsymbol{g})$ plays the role of a wrench observer in $(\mathcal{M},\tilde{\boldsymbol{g}})$. Finally, applying $\tilde{\boldsymbol{g}}^{-1}$ to the above quantities, there follow the \textit{transformed} vector fields
 \begin{equation}
\tilde{\boldsymbol{E}}=-\alpha\boldsymbol{X}\qquad{\rm and}\qquad \tilde{\boldsymbol{B}}=-\beta\boldsymbol{X},
\end{equation}
implying that $\tilde{\boldsymbol{W}}=-\boldsymbol{X}$ i.e., the opposite of the wrench observer in $(\mathcal{M},\boldsymbol{g})$ plays the role of the wrench 3-vector in $(\mathcal{M},\tilde{\boldsymbol{g}})$. Furthermore, it is straightforward to show, using Eq.\ (\ref{TXW}), that
\begin{equation}
||\tilde{\boldsymbol{E}}||^{2}_{\tilde{g}}=||\boldsymbol{E}||^{2}_{g}=-\alpha^{2},\quad\quad ||\tilde{\boldsymbol{B}}||^{2}_{\tilde{g}}=||\boldsymbol{B}||^{2}_{g}=-\beta^{2},\quad\quad\tilde{\boldsymbol{g}}(\tilde{\boldsymbol{E}},\tilde{\boldsymbol{B}})={\boldsymbol{g}}(\boldsymbol{E},\boldsymbol{B})=-\alpha\beta.
\end{equation}
With these relations one easily shows that the invariants are indeed the same.

In order to conclude these algebraic considerations, we compute how the hydrodynamics quantities, as given by Eq.\ (\ref{T}), appear to the new wrench observer. A direct calculation substituting the above formula into Eqs.\ (\ref{rho}) and (\ref{pi}) give
\begin{eqnarray}
\tilde{\rho}=\rho,\quad\quad \tilde{p}=p,\quad\quad \tilde{\boldsymbol{q}}=\boldsymbol{q}=0,\quad\quad\tilde{\boldsymbol{\Pi}}=\boldsymbol{\Pi}-\frac{4}{3}\kappa_{g}\left(\boldsymbol{l}^{\flat_{g}}\otimes\boldsymbol{n}^{\flat_{g}}+\boldsymbol{n}^{\flat_{g}}\otimes\boldsymbol{l}^{\flat_{g}}\right).
\end{eqnarray}
The reader may find it helpful to get a feel for these seemingly uncommon relations by starting straight from $\tilde{\boldsymbol{T}}=-\kappa\boldsymbol{g}$ and projecting onto $\boldsymbol{W}$ in the usual way. Interestingly, only the anisotropic pressure transforms.

\subsection{Curvature}

In order to see how the curvature transforms, let $(\mathcal{M}, \tilde{\boldsymbol{g}})$ and $(\mathcal{M},\boldsymbol{g})$ be space-times endowed with affine connections compatible with their corresponding metrics . The Levi-Civita connections associated to each space-time are linked through
\begin{equation}
\tilde\Gamma^{a}_{\phantom a bc}-\Gamma^{a}_{\phantom a bc}\doteq K^{a}_{\phantom a bc}=\frac{1}{2}\tilde g^{ad}(\nabla_{c}\tilde g_{bd} + \nabla_{b}\tilde g_{cd} - \nabla_{d}\tilde g_{bc})
\end{equation}
and a similar result stands for the difference between the Riemann curvature tensors
\begin{equation}
\tilde R^{a}{}_{bcd}-R^{a}{}_{bcd}\doteq B^{a}{}_{bcd}= \nabla_d K^{a}_{bc} - \nabla_c K^{a}_{bd} +K^{a}_{ed}K^{e}_{bc} - K^{a}_{ec}K^{e}_{bd}.
\end{equation}

Even for simple maps between the metrics, as the disformal transformation, it is a rather long calculation to compute the components of the tensors $K^{a}_{bc}$ and $B^{a}{}_{bcd}$ explicitly. But, with the help of Eqs.\ (\ref{g_null_basis}) and (\ref{tilde_g_null_basis}), we can construct null tetrad bases for both metrics creating a correspondence between the NP invariants of each space-time. From Eqs.\ (\ref{g_null_basis}), we see that ${\mathcal B}=\{\boldsymbol{l},\boldsymbol{n},\boldsymbol{m},\bar{\boldsymbol{m}}\}$ constitutes a basis for $\boldsymbol{g}$ whose vectors are also principal null directions of $\boldsymbol{F}$. On the other hand, this induces another basis $\tilde{{\mathcal B}}=\{\tilde{\boldsymbol{l}},\tilde{\boldsymbol{n}},\tilde{\boldsymbol{m}},\bar{\tilde{\boldsymbol{m}}}\}$ for the disformal metric $\tilde{\boldsymbol{g}}$ satisfying the same algebraic properties, maintaining the orientation [see Eq.\ (\ref{orient_basis})]. The new vectors and metric-related one-forms are given by
\begin{equation}\label{bases_map}
\begin{array}{l}
\tilde{\boldsymbol{l}}\rightarrow-\boldsymbol{n},\quad \tilde{\boldsymbol{n}}\rightarrow\boldsymbol{l},\quad\tilde{\boldsymbol{m}}\rightarrow\boldsymbol{m}.\\
\tilde{\boldsymbol{l}}^{\flat_{\tilde g}} \rightarrow  \boldsymbol{n}^{\flat_{g}},\quad \tilde{\boldsymbol{n}}^{\flat_{\tilde g}} \rightarrow -\boldsymbol{l}^{\flat_{g}},\quad\tilde{\boldsymbol{m}}^{\flat_{\tilde g}}\rightarrow\boldsymbol{m^{\flat_{g}}}.
\end{array}
\end{equation}
Thus, a tedious calculation shows that the ten invariant scalars corresponding to the Ricci tensors of each metric are linked as follows\footnote{See Appendix (\ref{np-appendix}) for details.}\\
\begin{equation}
\label{map_ricci}
\begin{array}{lcl}
\tilde\Phi_{00}&=& \Phi_{22} + [(\bar\alpha +  3\beta - 2\tau + \delta)\nu+{\rm c.c.}],\\
\tilde\Phi_{01}&=& -\Phi_{12} + \frac{1}{2}(\bar\gamma -\gamma + 2\mu + \bar\mu +\Delta) (\bar\pi+\tau) + \frac{1}{2}(\bar\alpha - 2\tau + \bar\pi +\beta + \delta) (\bar\mu - \mu) + \frac{1}{2} \bar\nu (\bar\varrho-\varrho)\\
&& + \frac{1}{2} \bar\lambda (\bar\tau +\pi),\\[1ex]
\tilde\Phi_{02}&=& \Phi_{02}+ (3\bar\varepsilon  -  \varepsilon  - 2\bar\varrho + D)\bar\lambda +(3 \gamma - \bar\gamma - 2\mu  - \Delta)\sigma,\\[1ex]
\tilde\Phi_{12}&=& \Phi_{01}-\frac{1}{2}(2\bar\pi-\beta-\tau-\alpha)+\delta)(\bar\varrho - \varrho) -\frac{1}{2}(2\bar\varrho +\varrho + \varepsilon - \bar\varepsilon - D) (\bar\pi+\tau)\\
&&- \frac{1}{2} (\mu-\bar\mu) \kappa - \frac{1}{2} (\bar\tau+\pi) \sigma ,\\[1ex]
\tilde\Phi_{11}&=& 6\Lambda +\frac{1}{4}[(D + \varepsilon+\bar\varepsilon-\bar\varrho - 2 \varrho)\mu - (\delta +\beta-\bar\alpha -2\tau)\pi  + (\bar\delta +\bar\beta - \alpha + 2\pi)\tau\\
&& - (\Delta + 2\mu -\bar\mu -\gamma-\bar\gamma )\varrho + {\rm c.c.}],\\
\tilde\Phi_{22}&=& \Phi_{00} + [(3 \alpha + \bar\beta - 2\pi - \bar\delta) \kappa +{\rm c.c.}],\\
\tilde\Lambda&=& \frac{1}{3}\Phi_{11}+\frac{1}{12}[(D + \varepsilon+\bar\varepsilon-\bar\varrho - 2 \varrho)\mu + (\delta + \beta-\bar\alpha -2\tau)\pi  - (\bar\delta +\bar\beta - \alpha + 2\pi)\tau\\
&& - (\Delta + 2\mu -\bar\mu-\gamma-\bar\gamma)\varrho + {\rm c.c.}],
\end{array}
\end{equation}
where ``c.c.'' means complex conjugation of terms that are between squared brackets. In particular, note that $\tilde\Phi_{00}$, $\tilde\Phi_{11}$, $\tilde\Phi_{22}$ and $\tilde\Lambda$ are real scalars, as expected.

Similarly, the five complex scalars associated with the corresponding Weyl tensors transform as follows

\begin{equation}
\label{map_weyl}
\begin{array}{lcl}
\tilde\Psi_0 &=& -\bar{\Psi}_4 - 2[ (\Delta + 3 \bar\gamma + 2 \mu - \gamma)\bar\lambda - (\tau + \bar\pi)\bar\nu],\\[1ex]
\tilde\Psi_1&=& -\bar{\Psi}_3 - \frac{3}{2}(\bar\gamma - \gamma + \bar\mu + \mu + \Delta) (\bar\pi +\tau) - \frac{1}{2}(\bar\alpha -4\tau +  \bar\pi + \beta + \delta) (\bar\mu - \mu)\\[1ex]
&&-  \frac{5}{2} \bar\lambda (\bar\tau + \pi)  -  \frac{1}{2} \bar\nu (\varrho - \bar\varrho)\\[1ex]
\tilde\Psi_2&=& \bar{\Psi}_2 + 2\Lambda - \frac{2}{3}\Phi_{11}
+\frac{1}{6} [3(\bar\varrho +\varrho) - \varepsilon - \bar\varepsilon - D](\bar\mu +\mu) + (\bar\alpha - \beta - \delta)(\pi +\bar\tau) + 2 \bar\lambda \bar\sigma + \frac{4}{3} \bar\tau \tau\\[1ex]
&& - \frac{1}{6}[7(\gamma + \bar\gamma) - \mu - \bar\mu - 7\Delta](\varrho + \bar\varrho)+ \frac{1}{6}[2\bar
\varrho \mu + (3\bar\pi-\tau+\beta-\bar\alpha +\delta)(\bar\tau -\pi) + c.c.]\\[1ex]
\tilde\Psi_3&=& \bar{\Psi}_1 - \frac{3}{2}(\varepsilon -\bar\varepsilon -\varrho -\bar\varrho + D)(\pi +\bar\tau) - \frac{1}{2}  (4\pi -\bar\beta-\bar\tau -\alpha + \bar\delta)(\bar\varrho-\varrho) \\[1ex]
&&+ \frac{1}{2} \bar\kappa(\mu - \bar\mu) + \frac{5}{2} \bar\sigma (\tau +\bar\pi)\\[1ex]
\tilde\Psi_4&=& -\bar{\Psi}_0 + 2[(\pi+\bar\tau)\bar\kappa + (\varepsilon - 2 \varrho - 3 \bar\varepsilon  + D)\bar\sigma].
\end{array}
\end{equation}

Finally, one easily shows that the invariant scalars associated to the electromagnetic field are transformed according with the simple relations
\begin{equation}
\tilde\Phi_{0}=\bar{\Phi}_2,\qquad\tilde{\Phi}_{1}=\Phi_1,\quad{\rm and}\quad\Phi_{2}=-\bar{\Phi}_0.
\end{equation}
Notice that none of the above transformation rules took into account neither the Maxwell's equations nor the Einstein's field equations. If we did so, the transformations would be rather simplified since only $\Phi_1$ is nonzero and the Ricci scalars of ($\mathcal{M},\boldsymbol{g}$) can be written in terms of it, besides of some restrictions on the spin coefficients.

\section{New solutions from old ones}

In this section we explore some consequences of the disformal transformation in the framework of three specific examples. Although they hardly exhaust all possibilities, we hope that the reader might get a feel of the exotic geometrical/topological properties implied by the transformation.

\subsection{Electrostatic fields}

We start with a static point particle with charge $Q$ located at the origin of the coordinate system and let $\mathcal{M}\equiv\mathbb{R}^{1,3}/\{\gamma\}$, with $\gamma$ denoting the worldline of the particle. In spherical coordinates $x^{a}=(t,r,\theta,\varphi)$, the Minkowski metric reads as:
\begin{equation}\label{Minksph}
\boldsymbol{g}=dt\otimes dt-dr\otimes dr- r^{2}d\theta\otimes d\theta-r^{2}\mbox{sin}^{2}\theta d\varphi\otimes d\varphi.
\end{equation}
The corresponding electromagnetic 2-forms are given by
\begin{equation}\label{EMpoint}
\boldsymbol{F}=-\frac{Q}{r^{2}}dr\wedge dt,\quad\quad\quad \star_{g}\boldsymbol{F}=-Q\ \mbox{sin}\theta\ d\theta\wedge d\varphi,
\end{equation}
and there follows the invariant
\begin{equation}\label{Kappapoint}
\kappa_{g}=\frac{1}{2}\left(\frac{Q^{2}}{r^{4}}\right),
\end{equation}
which obviously coincides with the energy density of the configuration.

It is straightforward to see that
\begin{equation}
\label{nul_bas_flat}
\boldsymbol{l}=\frac{1}{\sqrt{2}}(1,1,0,0), \quad \boldsymbol{n}=\frac{1}{\sqrt{2}}(1,-1,0,0), \quad{\rm and}\quad \boldsymbol{m}=\frac{1}{\sqrt{2}\,r}(0,0,1, i/\sin\theta)
\end{equation}
can provide a null tetrad basis for the metric (\ref{Minksph}) in terms of the NP formalism. Thus, the non-zero spin coefficient are $
\mu=\varrho=-1/\sqrt{2}r$ and $\beta=-\alpha=(1/4r)\sqrt{2} \cot\theta$.

Now, since $\partial/\partial t$ and $\partial/\partial r$ constitute a wrench observer and a wrench 3-vector in $(\mathcal{M},\boldsymbol{g})$, the disformal metric may be easily obtained putting $\boldsymbol{X}^{\flat_{g}}=dt$ and $\boldsymbol{W}^{\flat_{g}}=-dr$ in Eq.\ (\ref{TXW}) and substituting the result into Eq.\ (\ref{disf}), yielding
\begin{equation}\label{disfcharge}
\tilde{\boldsymbol{g}}=-dt\otimes dt+dr\otimes dr- r^{2}d\theta\otimes d\theta-r^{2}\mbox{sin}^{2}\theta d\varphi\otimes d\varphi.
\end{equation}
This metric has signature $(1,3)$ and is remarkably similar to the flat metric in Eq.\ (\ref{Minksph}). However, such simple modification has drastic consequences for the geometry $(\mathcal{M},\tilde{\boldsymbol{g}})$. Indeed, since the roles of $t$ and $r$ have been changed, $\tilde{\boldsymbol{g}}$ is actually time-dependent: surfaces of constant \textit{time} are now the level sets of the coordinate $r$ (equipotentials). A null tetrad basis for this metric can be obtained from (\ref{nul_bas_flat}) through the map (\ref{bases_map}) and it is easy to show that, under such transformation, the spin coefficients remain the same, except for $\tilde\mu=-\mu$. This subtle modification leads to non-zero NP invariant scalars as one can see straightforwardly from Eqs.\ (\ref{map_ricci}) and (\ref{map_weyl}), yielding
\begin{equation}
\tilde \Lambda=-\frac{1}{6r^2},\quad\tilde\Phi_{11}=-\frac{1}{2r^2},\quad \tilde\Psi_{2}=\frac{1}{3r^2}.
\end{equation}
Following \cite{acevedo}, it is possible to check by using these scalars that the metric (\ref{disfcharge}) is Petrov type-D.

A direct calculation shows that $W^{a}$ is geodesic, vorticity-free, with expansion coefficient and shear tensor given, respectively, by
\begin{eqnarray}
&&\tilde{\theta}=\tilde{\nabla}_{a}W^{a}=\partial_{r}(\ln\sqrt{-\tilde{g}})=2r^{-1},\\
&&\tilde{\sigma}_{ab}=\tilde{h}_{(a}^{\phantom a\phantom a c}\tilde{h}_{b)}^{\phantom a d}\tilde{\nabla}_{c}W_{d}-\frac{\tilde{\theta}}{3}\tilde{h}_{ab}=\frac{1}{3}\,\mbox{diag}(2r^{-1},0,-r,-r\mbox{sin}^{2}\theta),
\end{eqnarray}
where the round brackets in the definition of $\tilde{\sigma}_{ab}$ mean symmetrization.

\subsection{Magnetostatic fields}

Now, consider an infinite straight wire carrying a constant current $I$ and let $\mathcal{M}\equiv\mathbb{R}^{1,3}/\{\sigma\}$, with $\sigma$ the worldsheet of the wire. In cylindrical coordinates, $x^{a}=(t,\rho,\varphi,z)$, we have
\begin{equation}
\boldsymbol{g}=dt\otimes dt-d\rho\otimes d\rho- \rho^{2}d\varphi\otimes d\varphi- dz\otimes dz.
\end{equation}
The corresponding solution is given by
\begin{equation}
\boldsymbol{F}=\frac{I}{\rho}d\rho\wedge dz,\quad\quad\quad \star_{g}\boldsymbol{F}=I d\varphi\wedge dt.
\end{equation}
Notice that $\partial/\partial t$ is the wrench observer, whereas the wrench 3-vector is provided by $\rho^{-1}\partial/\partial\varphi$.

Putting $\boldsymbol{X}^{\flat_{g}}=dt$ and $\boldsymbol{W}^{\flat_{g}}=-\rho d\varphi$ in Eq.\ (\ref{TXW}) and substituting the result into Eq.\ (\ref{disf}), the disformal metric now reads as
\begin{equation}\label{disfcurrent}
\tilde{\boldsymbol{g}}=-dt\otimes dt-d\rho\otimes d\rho+ \rho^{2}d\varphi\otimes d\varphi-dz\otimes dz.
\end{equation}
Curiously, the disformal transformation in this case leads to a zero-curvature metric. Notwithstanding, the interesting features of (\ref{disfcurrent}) lies on its non-trivial topology: it has a compact time coordinate and, therefore, it is not globally hyperbolic admitting closed time-like curves. The only non-zero kinematic quantity associated with the wrench observers in this case is the acceleration given by $\tilde{a}_{a}=-(1/\rho)\delta^{\rho}_a$.

\subsection{Reissner-Nordstr\"om fields }

The geometry around a static charged black-hole with mass $M$ and charge $Q$ is given by the Reissner-Nordstr\"om metric
\begin{equation}\label{rn_metric}
\boldsymbol{g}=\frac{f}{r^2}dt\otimes dt-\frac{r^2}{f}dr\otimes dr- r^{2}d\theta\otimes d\theta-r^{2}\mbox{sin}^{2}\theta d\varphi\otimes d\varphi,
\end{equation}
where $f(r)\equiv r^2-2Mr+Q^2/2$, and the electromagnetic 2-forms are again given by Eq. (\ref{EMpoint}). A null tetrad basis for the metric (\ref{rn_metric}) can be given by
\begin{equation}
\label{nul_bas_rn}
\boldsymbol{l}= \frac{1}{\sqrt{2f}\,r}(r^2,f,0,0), \qquad \boldsymbol{n}= \frac{1}{\sqrt{2f}\,r}(r^2,-f,0,0), \qquad \boldsymbol{m}=\frac{1}{\sqrt{2}\,r}(0,0,1,i/\sin\theta)
\end{equation}
In this case, the non-zero spin coefficients are
\begin{equation}
\beta=\frac{\sqrt{2} \cot\theta}{4r}=-\alpha, \quad \mu=-\frac{\sqrt{f}}{\sqrt{2}r^2}=\rho,\quad \gamma=\frac{\sqrt{2}(2Mr-Q^2)}{8r^2\sqrt{f}}=\varepsilon
\end{equation}
and the non-zero NP invariant scalars are
\begin{equation}\Phi_{11}=-\frac{Q^2}{4r^4},\quad \Psi_{2}=\frac{2Mr-Q^2}{2r^4}.
\end{equation}

By noticing that $\kappa_{g}$ is given by Eq.\ (\ref{Kappapoint}), we obtain
\begin{equation}\nonumber
\tilde{\boldsymbol{g}}=-\left(1-\frac{2M}{r}+\frac{Q^{2}}{2r^{2}}\right)dt\otimes dt+\left(1-\frac{2M}{r}+\frac{Q^{2}}{2r^{2}}\right)^{-1}dr\otimes dr- r^{2}d\theta\otimes d\theta-r^{2}\mbox{sin}^{2}\theta d\varphi\otimes d\varphi
\end{equation}
which is, again, a simple permutation of the signs of $g_{00}$ and $g_{11}$. Now, a null tetrad basis for this metric can be obtained from (\ref{nul_bas_rn}) also through the map (\ref{bases_map}) and, under such transformation, the spin coefficients remain the same, exceptions for $\tilde\mu=-\mu$ and $\tilde\gamma=-\gamma$. The non-zero NP invariant scalars are now
\begin{equation}
\tilde\Lambda=-\frac{1}{6r^2},\quad\tilde{\Phi}_{11}=\frac{Q^2-2r^2}{4r^4},\quad \tilde{\Psi}_{2}=\frac{2r^2-6Mr+3Q^2}{6r^4}.
\end{equation}
Notice that we can recover the previous case of electrostatic fields in the limit $M,Q\rightarrow0$.

\section{Conclusion}
We have scrutinized some algebraic and differential features of source-free electromagnetic fields under the action of disformal transformations. Although similar transformations were investigated before \cite{Harte, GT}, to the best of our knowledge, no previous efforts have been given to better understand their geometrical/topological implications. In particular, we have seen that the energy-momentum tensor of a regular electromagnetic field plays an important role in the definition of its disformally-related metric. Conversely, the disformal energy-momentum tensor is proportional to the physical metric. This poses an interesting question. Can we distinguish between the metrical properties of a space-time and the distribution of energy-momentum corresponding to a test field? Our results indicate that, as far as we are concerned solely with regular electromagnetic fields in vacuum, the answer is no! However, by focusing on electromagnetic wrench configurations, we have clarified that the disformal metric undergoes a curious signature transition with respect to the physical metric: time and space exchange their roles. We have shown that time in the disformal manifold runs in the direction of the field lines.

We have also calculated how the disformal transformation acts on all the NP invariant scalars, leading to complicate expressions in terms of the spin coefficients only and mixing the scalars. Then, we apply those formulas to three well-known solutions of Maxwell's theory in order to see, in practice, the effect of the transformation on the disformal space-time. It turns out that two out of the three scenarios we have analyzed, the disformal transformation led to completely different space-times, although their metric components solely differ by an exchange of sign between two metric components. On the other hand, in the magnetostatic case, all the NP invariants are null in both metrics, while the actual change occurs in the topology. In particular, the disformal manifold has a compact time coordinate, allowing for closed time-like curves.

At last, we wonder whether these new (disformal) metrics might be taken indeed as space-time geometries of real physical systems. As seen along the paper, they do not share the causal structure of their corresponding physical manifold and the symmetries inherited from it led to space-time singularities or non-trivial topology. In spite of this, they seem to describe well-posed electromagnetic problems provided that the outcome should be reinterpreted accordingly. For instance, the electrostatic case in the physical manifold became a time-dependent electric field in an asymptotically flat disformal metric. Other interesting features to explore in these scenarios is particle physics, quantization approaches and connections with Einstein's field equations. We shall investigate these issues in a forthcoming communication.

\appendix

\section{Notation}\label{notation}

The signature convention $(1,3)$ means $+---$. At a space-time point, the set of all real $2$-forms is denoted by $\Lambda^{2}(T_{p}^{*}\mathcal{M})$, the second exterior power of the cotangent space. In components, the Hodge dual of a 2-form $\boldsymbol{G}\in\Lambda^{2}(T_{p}^{*}\mathcal{M})$ reads as
 \begin{equation}\label{Inner}
\star_{g}G_{ab}=\frac{1}{2}\sqrt{-g}[abcd](g^{-1})^{cp}(g^{-1})^{dq}G_{pq},
\end{equation}
where $g\equiv\mbox{det}(g_{ab})$, $[abcd]$ is the Levi-Civita permutation symbol with $[0123]=1$ and $(g^{-1})^{pc}$ are the contravariant components of the metric. For arbitrary 2-forms $\boldsymbol{G},\boldsymbol{H}$, the inner product induced by $\boldsymbol{g}$ on $\Lambda^{2}(T_{p}^{*}\mathcal{M})$ read as
\begin{equation}\label{innerGH}
\langle \boldsymbol{G},\boldsymbol{H}\rangle_{g}\equiv \frac{1}{2}(g^{-1})^{ac}(g^{-1})^{bd}G_{ab}H_{cd}
\end{equation}
and we shall stick to the following conventions
\begin{eqnarray*}
&&[(\boldsymbol{A}\boldsymbol{B})_{h}]_{ab}\equiv (h^{-1})^{cd}A_{ac}B_{db},\quad\quad\quad \mbox{tr}_{h}(\boldsymbol{A})\equiv (h^{-1})^{ab}A_{ab},\\\\
&&\ \ \ \boldsymbol{A}^{l_{h}}\equiv A_{a_{1}b_{1}}(h^{-1})^{b_{1}a_{2}}A_{a_{2}b_{2}}\ ...\ A_{a_{l-1}b_{l-1}}(h^{-1})^{b_{l-1}a_{l}}A_{a_{l}b_{l}},
\end{eqnarray*}
where $\boldsymbol{A},\boldsymbol{B},\boldsymbol{h}$ are $(0,2)$-tensors with  $\boldsymbol{h}$ non-degenerate and $l\in\mathbb{N}$. As usual, if $\boldsymbol{X}\in T_{p}\mathcal{M}$, the symbol $X^{\flat_{g}}\in T_{p}^{*}\mathcal{M}$ represents the musical isomorphism induced by $\boldsymbol{g}$ i.e., $\boldsymbol{X}^{\flat_{g}}=\boldsymbol{g}(\ \cdot\ ,\boldsymbol{X})$. The interior multiplication $i_\textbf{X}\textbf{F}$ is obtained from the 2-form by inserting $\boldsymbol{X}$ into the first slot i.e.,
\begin{equation}
(i_\textbf{X}\textbf{F})_{a}=X^{b}F_{ba}=-X^{b}F_{ab},
\end{equation}
whereas the orthogonal complement induced by $\boldsymbol{g}$ is denoted by $\boldsymbol{X}^{\perp_{g}}$.

\section{Newman-Penrose formalism}\label{np-appendix}
Following \cite{chandra}, let us introduce a complex null tetrad basis $\{\boldsymbol{l},\boldsymbol{n},\boldsymbol{m},\bar {\boldsymbol{m}}\}$ satisfying the following normalization conditions
\begin{eqnarray}
&&\boldsymbol{l}\cdot\boldsymbol{l}=0,\quad\quad \boldsymbol{n}\cdot\boldsymbol{n}=0,\quad\quad\boldsymbol{m}\cdot\boldsymbol{m}=0,\quad\quad \bar{\boldsymbol{m}}\cdot\bar{\boldsymbol{m}}=0,
\end{eqnarray}
and with all possible cross-relations zero with the exception of
\begin{equation}
\boldsymbol{l}\cdot\boldsymbol{n}=1,\quad\quad \boldsymbol{m}\cdot\bar{\boldsymbol{m}}=-1.
\end{equation}
Our orientation convention is such that
\begin{equation}
\boldsymbol{\varepsilon}(\boldsymbol{e}_{0},\boldsymbol{e}_{1},\boldsymbol{e}_{2},\boldsymbol{e}_{3})=+1
\end{equation}
whenever $\{\boldsymbol{e}_{0},\boldsymbol{e}_{1},\boldsymbol{e}_{2},\boldsymbol{e}_{3}\}$ is a pseudo-orthonormal basis with $\boldsymbol{e}_{0}$ future-directed and the space-like vectors forming a right-handed triad. This implies in
\begin{equation}
\label{orient_basis}
\varepsilon_{abcd}l^{a}n^{b}m^{c}\bar{m}^{d}=i
\end{equation}
with $\boldsymbol{l}$ and $\boldsymbol{m}$ also future-directed. The directional derivatives are defined by
$$D\equiv l^a\nabla_a,\quad \Delta\equiv n^a\nabla_a,\quad \delta\equiv m^a\nabla_a,\quad \bar{\delta}\equiv \bar m^a\nabla_a,$$
while the transportation equations read as follows
\begin{equation}
\begin{array}{lcl}
D l^a&=&(\epsilon + \bar\epsilon)l^a - \bar \kappa m^a - \kappa \bar m^a,\\
\Delta l^a&=&(\gamma + \bar\gamma)l^a - \bar \tau m^a - \tau \bar m^a,\\
\delta l^a&=&(\beta + \bar\alpha)l^a - \bar \rho m^a - \sigma \bar m^a,\\
\bar \delta l^a&=&(\alpha + \bar\beta)l^a - \bar \sigma m^a - \rho \bar m^a,\\
D n^a&=&-(\epsilon + \bar\epsilon)n^a + \pi m^a + \bar\pi \bar m^a,\\
\Delta n^a&=& -(\gamma + \bar\gamma)n^a + \nu m^a + \bar\nu \bar m^a,\\
\delta n^a&=&-(\beta + \bar\alpha)n^a + \mu m^a + \bar\lambda \bar m^a,\\
\bar \delta n^a&=&-(\alpha + \bar\beta)n^a + \lambda m^a + \bar\mu \bar m^a,\\
D m^a&=&(\epsilon - \bar\epsilon)m^a + \bar \pi l^a - \kappa n^a,\\
\Delta m^a&=&(\gamma - \bar\gamma)m^a + \bar \nu l^a - \tau n^a,\\
\delta m^a&=&(\beta - \bar\alpha)m^a + \bar \lambda l^a - \sigma n^a,\\
\bar \delta m^a&=&(\alpha - \bar\beta)m^a + \bar \mu l^a - \rho n^a,\\
D \bar m^a&=&(\bar\epsilon - \epsilon)\bar m^a + \pi l^a - \bar \kappa n^a,\\
\Delta \bar m^a&=&(\bar\gamma - \gamma)\bar m^a + \nu l^a - \bar \tau n^a,\\
\delta \bar m^a&=&(\bar\alpha - \beta)\bar m^a + \mu l^a - \bar \rho n^a,\\
\bar \delta \bar m^a&=& (\bar\beta - \alpha)\bar m^a + \lambda l^a - \bar\sigma n^a.
\end{array}
\end{equation}

The Ricci invariants are defined as
\begin{eqnarray}
\Phi_{00}&=& - \frac{1}{2}R_{ab}\,l^a l^b,\\
\Phi_{01}&=& - \frac{1}{2}R_{ab}\,l^a m^b,\\
\Phi_{02}&=& - \frac{1}{2}R_{ab}\,m^a m^b,\\
\Phi_{12}&=& - \frac{1}{2}R_{ab}\,m^a n^b,\\
\Phi_{11}&=& - \frac{1}{4}R_{ab}\,(l^a n^b + m^a \bar m^b),\\[1ex]
\Phi_{22}&=& - \frac{1}{2}R_{ab}\,n^a n^b,\\
\Lambda&=& \frac{1}{24}R=\frac{1}{12}R_{ab}\,(l^a n^b - m^a \bar m^b),
\end{eqnarray}
whereas the Weyl invariants read as follows
\begin{equation}
\begin{array}{lcl}
\Psi_0&=&-C_{abcd}\,l^a m^b l^c m^d,\\
\Psi_1&=&-C_{abcd}\,l^a n^b l^c m^d,\\
\Psi_2&=&-C_{abcd}\,l^a m^b \bar m^c n^d,\\
\Psi_3&=&-C_{abcd}\,l^a n^b \bar m^c m^d,\\
\Psi_4&=&-C_{abcd}\,\bar m^a n^b \bar m^c n^d.
\end{array}
\end{equation}

\end{document}